\documentclass[10pt]{iopart}
 \usepackage[numbers,sort&compress]{natbib}
\usepackage{graphicx}

\usepackage{natbib}
\begin{document}

\title{Design of infrared microspectrometers based on phase-modulated axilenses}

\author{Yuyao Chen$^1$, Wesley A. Britton$^2$, and Luca  Dal Negro$^{1,2,3}$}

\address{$^1$ Department of Electrical \& Computer Engineering and Photonics Center, Boston University, 8 Saint Mary’s Street, Boston, MA 02215, USA}
\address{$^2$ Division of Material Science and Engineering, Boston University, 15 Saint Mary’s Street, Brookline, MA 02446, USA}
\address{$^3$ Department of Physics, Boston University, 590 Commonwealth Avenue, Boston, MA 02215, USA}
\ead{dalnegro@bu.edu}

\begin{abstract}
We design and characterize a novel axilens-based diffractive optics platform that flexibly combines efficient point focusing and grating selectivity and is compatible with scalable top-down fabrication based on a 4-level phase mask configuration. This is achieved using phase-modulated compact axilens devices that simultaneously focus incident radiation of selected wavelengths at predefined locations with larger focal depths compared to traditional Fresnel lenses. In addition, the proposed devices are polarization insensitive and maintain a large focusing efficiency over a broad spectral band. Specifically, here we discuss and characterize modulated axilens configurations designed for long-wavelength infrared (LWIR) in the $6~\mu$m--12~$\mu$m wavelength range and in the $4~\mu$m--6~$\mu$m mid-wavelength infrared (MWIR) range. These devices are ideally suited for monolithic integration atop the substrate layers of infrared focal plane arrays (IR-FPAs) and for use as compact microspectrometers. We systematically study their focusing efficiency, spectral response, and cross talk ratio, and we demonstrate linear control of multi-wavelength focusing on a single plane. Our design method leverages Rayleigh-Sommerfeld  (RS) diffraction theory and is validated numerically using the Finite Element Method (FEM). Finally, we demonstrate the application of spatially modulated axilenses to the realization of compact, single-lens spectrometer. By optimizing our devices, we achieve a minimum distinguishable wavelength interval of $\Delta\lambda=240nm$ at $\lambda_c=8{\mu}m$ and $\Delta\lambda=165nm$ at $\lambda_c=5{\mu}m$. The proposed devices add fundamental spectroscopic capabilities to compact imaging devices for a number of applications ranging from spectral sorting to LWIR and MWIR phase contrast imaging and detection. 
\end{abstract}

%
%
%
%
\maketitle
\ioptwocol
\section{Introduction}
Infrared focal plane arrays (IR-FPAs) are at the heart of present IR imaging and sensing technology \cite{krishna2007quantum,stiff2009quantum,choi2017resonant,zhang2018solid}. In order to increase the sensitivity and suppress the crosstalk, microlenses have been proposed as concentrators of radiation in combination with IR-FPAs \cite{chen2002monolithic,jian2012design,bai2014performance,akin2015mid,allen2016increasing}. Recently, metasurfaces have received significant attention due to their compact size and optically flat profiles \cite{arbabi2015subwavelength,khorasaninejad2016polarization,chen2019broadband}. However, these devices rely either on engineered resonance behavior, which introduces unavoidable losses and reduce the overall focusing efficiency \cite{zhan2016low,khorasaninejad2016polarization}, or on geometrical phase modulation that requires polarization control \cite{lin2014dielectric,zheng2015metasurface}. These problems can be overcome through advanced metasurface designs that require precise sub-wavelength accuracy and add to costs and device complexity. State-of-the-art achromatic metalens-based devices \cite{arbabi2016multiwavelength,chen2018broadband,shrestha2018broadband} and super-oscillation lenses \cite{yuan2017achromatic} recently achieved outstanding performances but at further increased fabrication demands.

In this paper, we propose an alternative microlens design approach based on the phase modulation of compact axilenses  that can be monolithically integrated atop the substrate of IR-FPAs. These devices efficiently combine the focusing behavior of an axilens \cite{davidson1991holographic} with the spatial dispersion behavior of a grating structure. In our work, we discretize the phase profile of the resulting devices in order to create novel 4-level diffractive optical elements (DOEs) that are compatible with top-down photolithography. Notably, the proposed structures are polarization insensitive and can be readily designed for applications to any desired spectral range. Consequently, the proposed concepts enable the engineering of compact, on-chip diffractive focusing devices for a number of applications to photodetection, on-chip spectroscopy, and imaging.
In what follows, we study the wave diffraction problem in a nmber of realizations of different 4-level modulated axilenses using the rigorous Rayleigh-Sommerfeld (RS) first integral formulation \cite{goodman2005introduction}. First, we demonstrate that axilenses are capable of focusing radiation of different wavelengths on the same plane. The existence of an achromatic plane enables angular dispersion effects and wavelength selective beam steering across the focal plane when additional transverse phase modulation are added. In order to prove this point, we consider one-dimensional (1D) and two-dimensional (2D) periodic as well as chirp phase modulations within a 4-level mask geometry with a maximum diameter $D=100{\mu}m$, which is a size comparable with typical metalens approaches \cite{arbabi2015subwavelength,chen2018broadband}. Thirdly, we show that engineered phase-modulated axilenses can tightly focus radiation of multiple wavelengths on the same plane at different predefined locations, enabling compact spectrometer for LWIR and MWIR.

\section{Achromatic focusing of axilenses}
Before introducing our novel designs, we review the basic operation of axilenses, which are characterized by a phase profile given by:
\begin{equation}\label{phase profile of axilens}
\left.\phi(r)\right\vert_{2\pi}=-\frac{2\pi}{\lambda}\left[\sqrt{\left(f_0+\frac{r{\Delta}f}{R}\right)^2+r^2}-\left(f_0+\frac{r{\Delta}f}{R}\right)\right]
\end{equation}
where $r=\sqrt{x^2+y^2}$, $f_0$ is the focal length, and ${\Delta}f$ is the focal depth. The $2\pi$ subscript indicates that the phase is reduced by modulo $2\pi$. Differently from a traditional Fresnel lens, an axilens can focus incident radiation with a focal depth that can be controlled by changing ${\Delta}f$. 

\begin{figure}[h!]
\centering
\includegraphics[width=\linewidth]{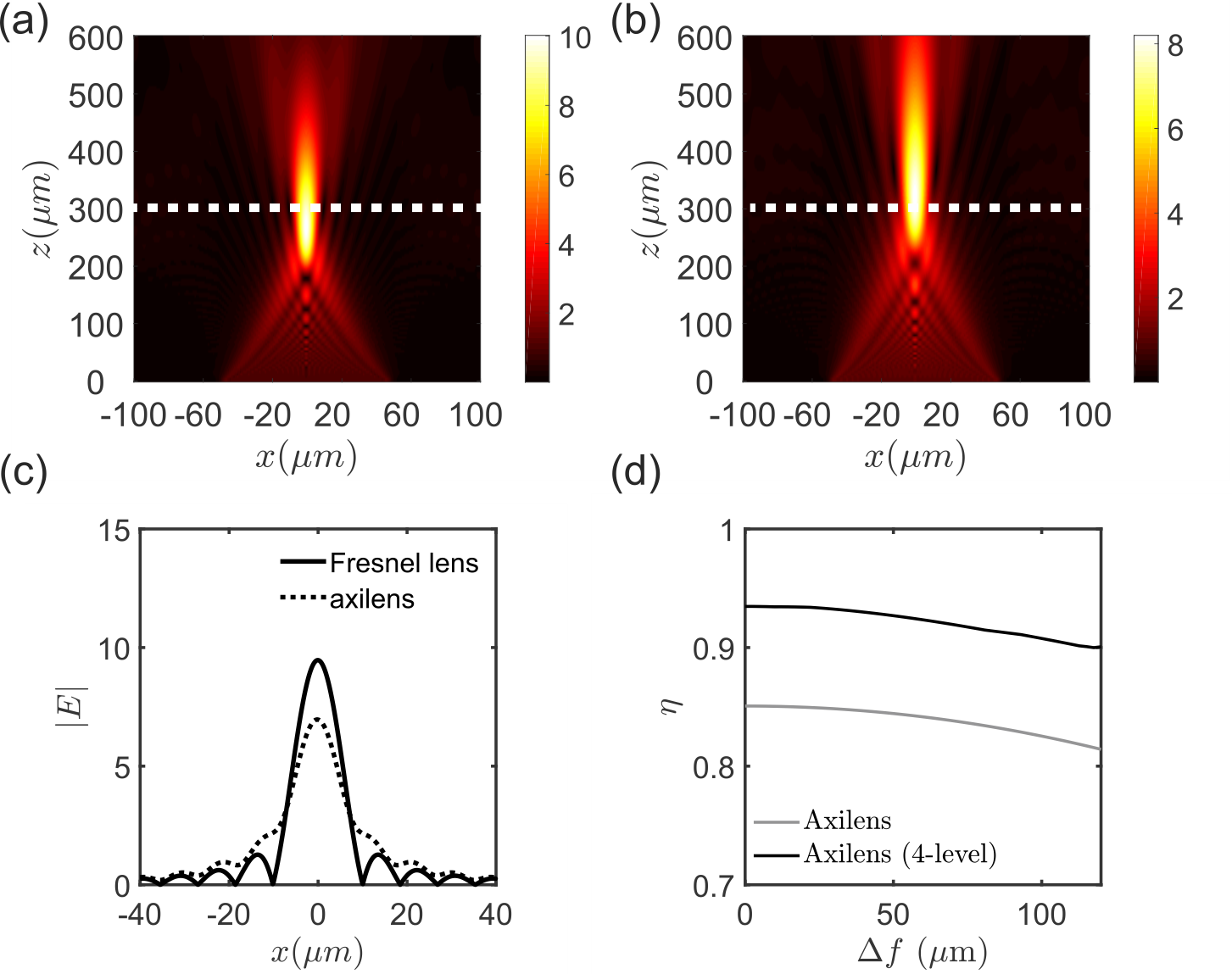}
\caption{(a) The focusing side view of a standard Fresnel lens with focal length $f_0=300{\mu}m$. (b) The focusing side view of an axilens with $f_0=250{\mu}m$ and ${\Delta}f=120{\mu}m$. (c) displays the field amplitude cut along $z=300{\mu}m$ in (a, b). (d) shows the focusing efficiency of an axilens with respect to ${\Delta}f$ with $f_0=300{\mu}m$ and the incident wavelength set to $\lambda=9{\mu}m$ in the substrate.}
\label{Fig1}
\end{figure}

We consider devices located on the $xy$ plane and excited by a normally incident plane wave giving rise to a diffracted field that propagates along the $+z$ direction. In Fig. \ref{Fig1} (a) and (b), we compare the computed field amplitude side view (side view on the $xz$ plane) for a standard Fresnel lens and an axilens computed according to the RS with the following expressions \cite{goodman2005introduction}:
\begin{eqnarray}\label{RS equation}
    U_{2}\left(x^\prime,y^\prime\right)=U_{1}\left(x,y\right) * h(x,y)\\
    h(x,y)=\frac{1}{2\pi}\frac{z}{r}  \left(\frac{1}{r}-jk\right)\frac{e^{\left(jkr\right)}}{r}
\end{eqnarray}
where $*$ denotes 2D space convolution, $U_1$, $U_2$ are the transverse field distributions in the object and image planes with coordinates $(x,y)$ and $(x^\prime,y^\prime)$, respectively. Moreover, $k$ is the incident wave number and $r=\sqrt{x^2+y^2+z^2}$, where $z$ is the distance between object and image plane.

\begin{figure}[h!]
\centering
\includegraphics[width=\linewidth]{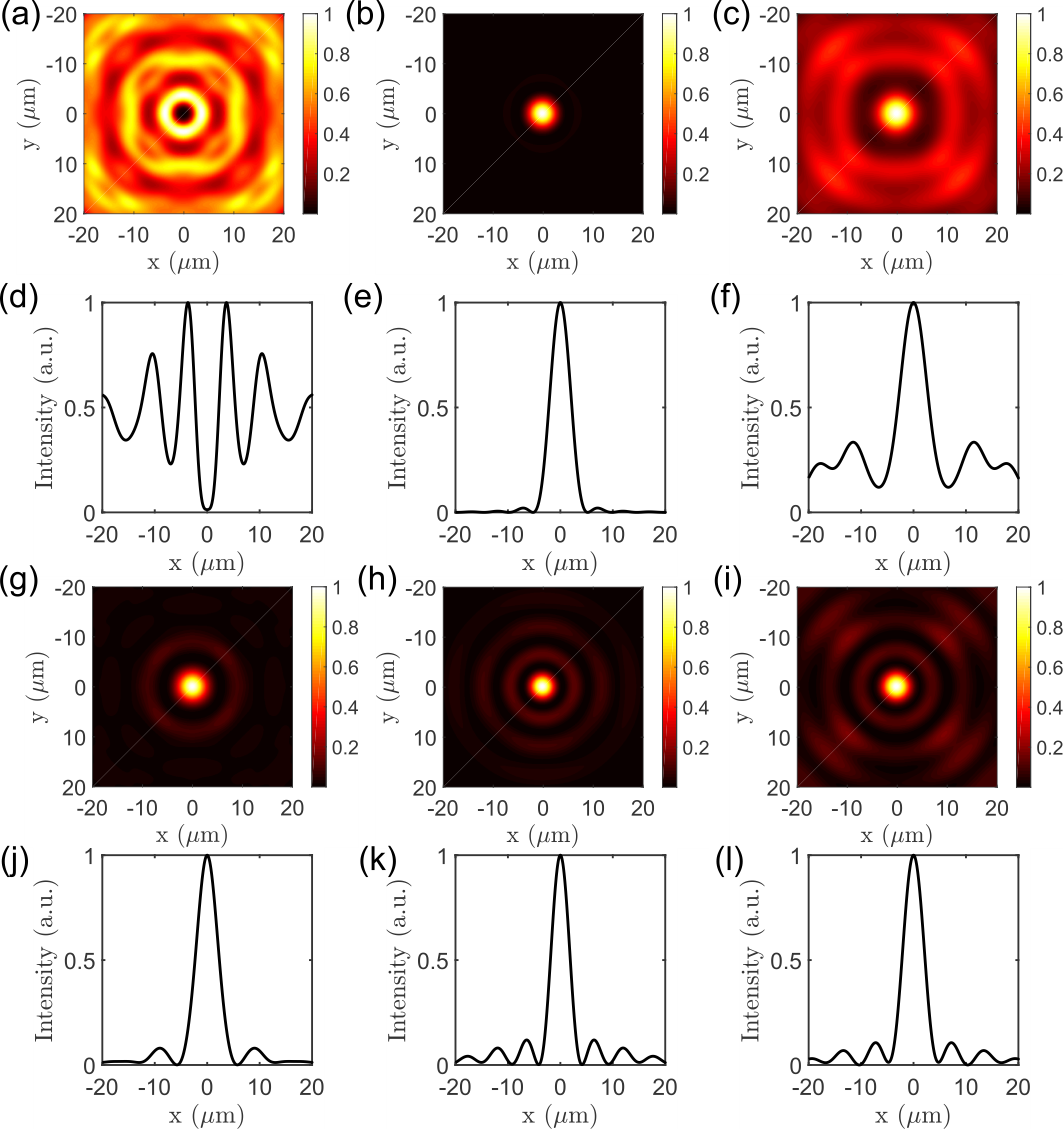}
\caption{(a-c) show the normalized 2D intensity profiles of a Fresnel lens with $f_0=300{\mu}m$ at the focal plane $z=300{\mu}m$ with incident wavelengths $\lambda=6  {\mu}m, 9 {\mu}m, 12 {\mu}m$, respectively. (d-f) show the corresponding 1D intensity cut profiles corresponding to panels (a-c). Panels (g-f) show the normalized 2D intensity profiles of Fresnel lens with $f_0=250{\mu}m$, ${\Delta}f=120 {\mu}m$ at the focal plane $z=300{\mu}m$ with incident wavelengths $\lambda=6  {\mu}m, 9 {\mu}m, 12 {\mu}m$, respectively. Panels (j-l) show the 1D intensity cut profiles corresponding to panels (g-f).}
\label{Fig2}
\end{figure}

We design both the Fresnel lens and the axilens in order to focus radiation of wavelength $\lambda=9{\mu}m$ at $z=300{\mu}m$. We consider the presence of a substrate with refractive index $n=3.3$, as typically used in IR-FPAs platforms, and scale the wavelength accordingly. In Fig. \ref{Fig1} (c) we show a transverse cut along the $x$-axis of the field amplitude at the detection plane ($z=300{\mu}m$), highlighted by the dashed lines in panels (a) and (b). Our results show that although the axilens features a broader focal spot in the transverse plane, it can focus incident radiation over a significantly longer longitudinal range along the $z$-axis. In fact, a desired focal depth ${\Delta}f$ can be obtained using an axilens regardless of the wavelength of the incident radiation \cite{vijayakumar2017design}. Although different wavelengths are focused at different positions along the $z$-axis, the larger focal depth of axilenses compared to traditional lenses guarantees a substantial overlap of the focused intensities of different wavelengths on the same plane, thus establishing an achromatic focal plane. This important feature of the axilenses will be explored later in the paper to demonstrate single-lens spectrometers based on phase-modulated axilenses. However, in practical applications it is important to quantify the focusing efficiency $\eta$ of such devices. This is done by defining the quantity $\eta$ as the ratio, with respect to the total incident power, of the power confined within a circular region with a radius that is three times larger than the full-width-at-half-maximum (FWHM) of the focal spot.  For a better comparison, we consider an axilens design with $f_0=300{\mu}m$ and we evaluate $\eta$ as a function of ${\Delta}f$, where ${\Delta}f=0$ corresponds to a standard Fresnel lens. In Fig. 1 (d) we display the results of our analysis. We compare devices with continuous phase distribution and with 4-level discretized phase \cite{pawlowski1997multilevel}.  The analysis indicates that increasing $\Delta f$ will decrease the focusing efficiency due to the broader focal spot as shown in Fig. \ref{Fig1} (c). However, a 4-level discretized axilens maintains a large $\eta$ $(\sim 80\%)$ over a broad ${\Delta}f$ range and gives rise to achromatic focusing over a broad spectral band as shown in Fig. \ref{Fig2}.

 In Fig. \ref{Fig2} we directly demonstrate the distinctive achromatic focusing behavior of the axilens compared to traditional Fresnel lens across the LWIR spectrum. In particular, we focus at wavelengths  $\lambda=6{\mu}m$, $9{\mu}m$, and $12{\mu}m$, and compute the 2D intensity profiles at the focal plane for a Fresnel lens (Figs. \ref{Fig2} (a-c)) and a axilens (Figs. \ref{Fig2} (g-i)). The corresponding 1D horizontal cuts along the centers are shown in Figs. \ref{Fig2} (d-f) for the Fresnel lens and Figs. \ref{Fig2} (j-l) for the axilens. The achromatic behavior of the axilens across the LWIR spectrum is clearly manifested by the almost constant width of the focusing point-spread functions (along the transverse $x$-axis) shown in Figs. \ref{Fig2} (j-l) . 
 
\section{Phase-modulated axilenses}
\begin{figure}[h!]
\centering
\includegraphics[width=\linewidth]{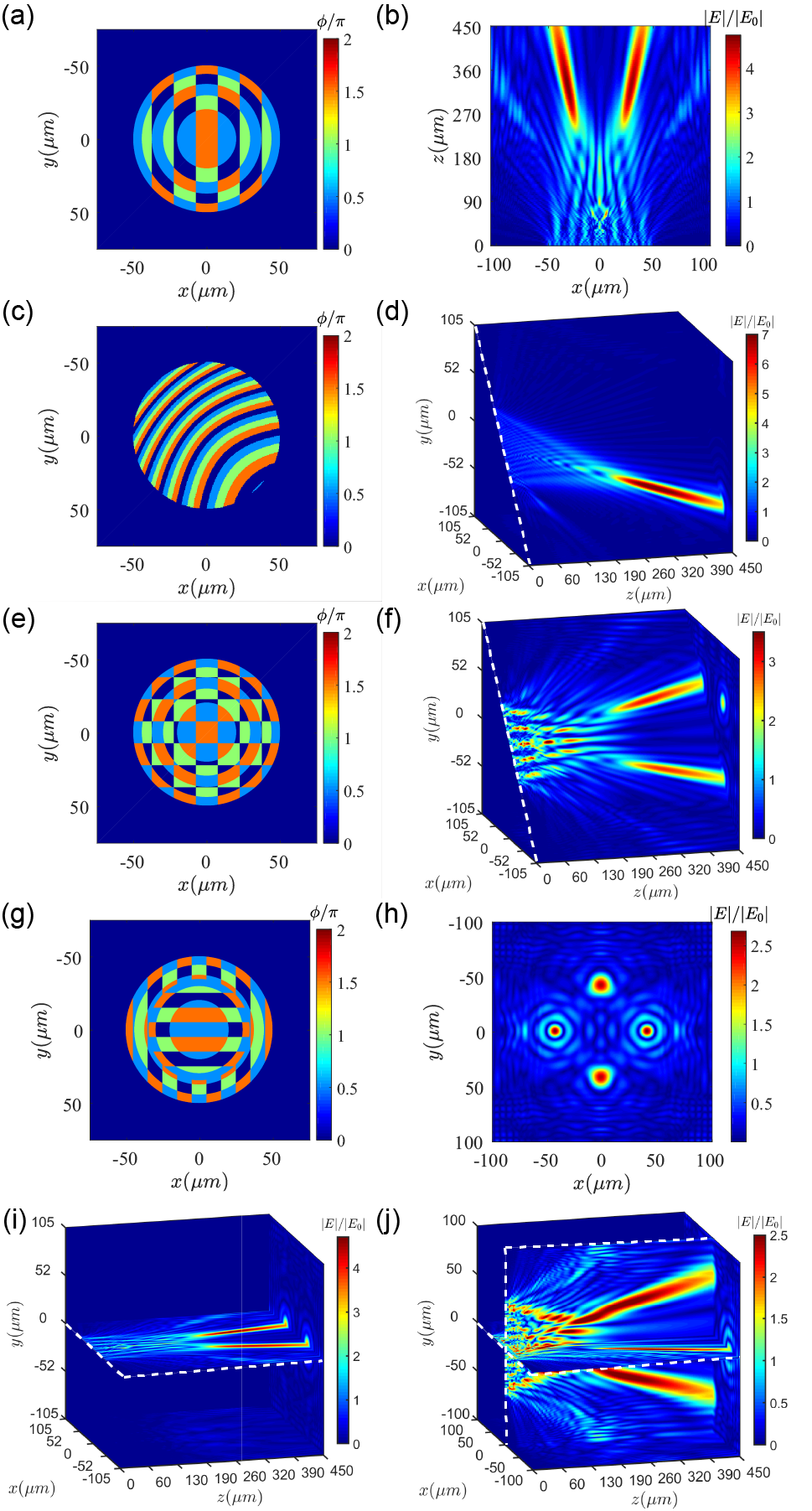}
\caption{Panels (a,c,e,g) show the 4-level discretized phase profiles for 1D periodically modulated axilens, a $45^\circ$ chirped axilens, a 2D periodically modulated axilens, and a cross modulated axilens, respectively. Panel (b) shows the simulated diffraction field side view of the 1D modulated axilens. Panels (d,f) show the diffraction field 3D view for the phase profile in (c) and (e), respectively. Panel (h) shows the simulated field amplitude distribution at the focal plane $z=300{\mu}m$. Panels (i,j) display the diffraction field 3D view for the phase profiles shown in (a) and (g), respectively.}
\label{Fig3}
\end{figure}

We now consider the effects of periodic and chirped phase modulations in 4-level discretized axilens devices. Phase-modulated axilenses are examples extending the reach of multifunctional DOEs, which are used for beam shaping applications \cite{vasara1989realization,vijayakumar2015conical}. In Figs. \ref{Fig3} (a-c), we show the phase profiles of modulated axilenses summed (modulo 2$\pi$) with a 1D periodic function and a $45^\circ$ chirped function, respectively. The phase profile parameters of the axilens were chosen as follows: $f_0=250{\mu}m$, ${\Delta}f=120{\mu}m$, ${\lambda}=9{\mu}m$, $n=3.3$. These parameters can be flexibly changed to address other photodetector spectral ranges and/or dielectric substrate thickness and refractive indices. The phase of the axilenses was modulated by adding a periodic phase with period equal to $p=15{\mu}m$ as well as a chirping function with the position-dependent phase profile $2\pi (x^\prime+y^\prime)/p$, $p=15{\mu}m$. We also show the phase profile of a 2D periodically modulated axilens in Fig. \ref{Fig3} (e) with periodicity $p=15{\mu}m$. Finally, figure \ref{Fig3} (g) depicts an alternative implementation of the periodically modulated axilens where two radially interlocked periodic functions were used with periodicity $p=10{\mu}m$ along vertical and horizontal directions inside two separate concentric regions of the device (annular crossed modulations). The ratio between the radii of the outer and inner region of the device is $1.455$ and was optimized to guarantee the equal intensity at the four focal spots. All the modulated axilenses have a diameter $D=100{\mu}m$, which is comparable to the typical size of alternative devices based on metalenses \cite{chen2018broadband,zhang2018solid}. Figure \ref{Fig3} (b) shows the horizontal $xz$-plane side view of the simulated diffraction field amplitude normalized by incident plane wave amplitude $E_0$ of the periodically modulated axilens corresponding to Fig. \ref{Fig3} (a). We find that the incoming radiation is split into two focused beam components. This is due to the spatial dispersion of the ${\pm}1$ diffraction orders associated to the periodic phase modulation while each beam is focused with a larger depth of focusing due to the characteristic focusing phase profile of the axilens. Therefore, the periodically modulated axilens is capable of simultaneously splitting and focusing different IR bands at different locations along the transverse $x$-axis. This effect can be designed to match the different pixel locations in a realistic IR-FPAs implementation. We further show in Fig \ref{Fig3} (d) the three-dimensional (3D) distribution along a plane cut at $45^\circ$ of the diffracted field that is focused by the device with phase modulation shown in Fig \ref{Fig3} (c). In this case, radiation of different wavelengths is focused at different points along the diagonal ($45^\circ$) line that cuts through the focusing plane, further demonstrating control of the focal spot's trajectory. Figure \ref{Fig3} (f) illustrates the 3D distribution of the diffracted field of the 2D modulated axilens whose phase is shown in Fig. \ref{Fig3} (e). This device focuses incident radiation into four focal spots arranged along the diagonal directions, where the 45$^\circ$ plane cut in Fig. \ref{Fig3} (f) shows two focal spots along one diagonal direction. Finally, in Fig. \ref{Fig3} (h) we show the diffracted field amplitude distribution across the transverse $xy$ focal plane for the device with the phase shown in Fig. \ref{Fig3} (g). In this case, the radiation is diffracted into four focal spots along the $y$- and $x$-axis due to the phase mataching at the inner region and the outer region of the device, respectively. Figure \ref{Fig3} (i) and (j) show the 3D view of diffraction fields of devices with phase profiles shown in panels (a) and (g), respectively. The diffraction field are sampled on the $xz$-plane and both $xz$- and $yz$-plane for panels (i) and (j), respectively. Notice that all the devices have been designed to focus at the same plane $z=300{\mu}m$.

\begin{figure}[h!]
\centering
\includegraphics[width=\linewidth]{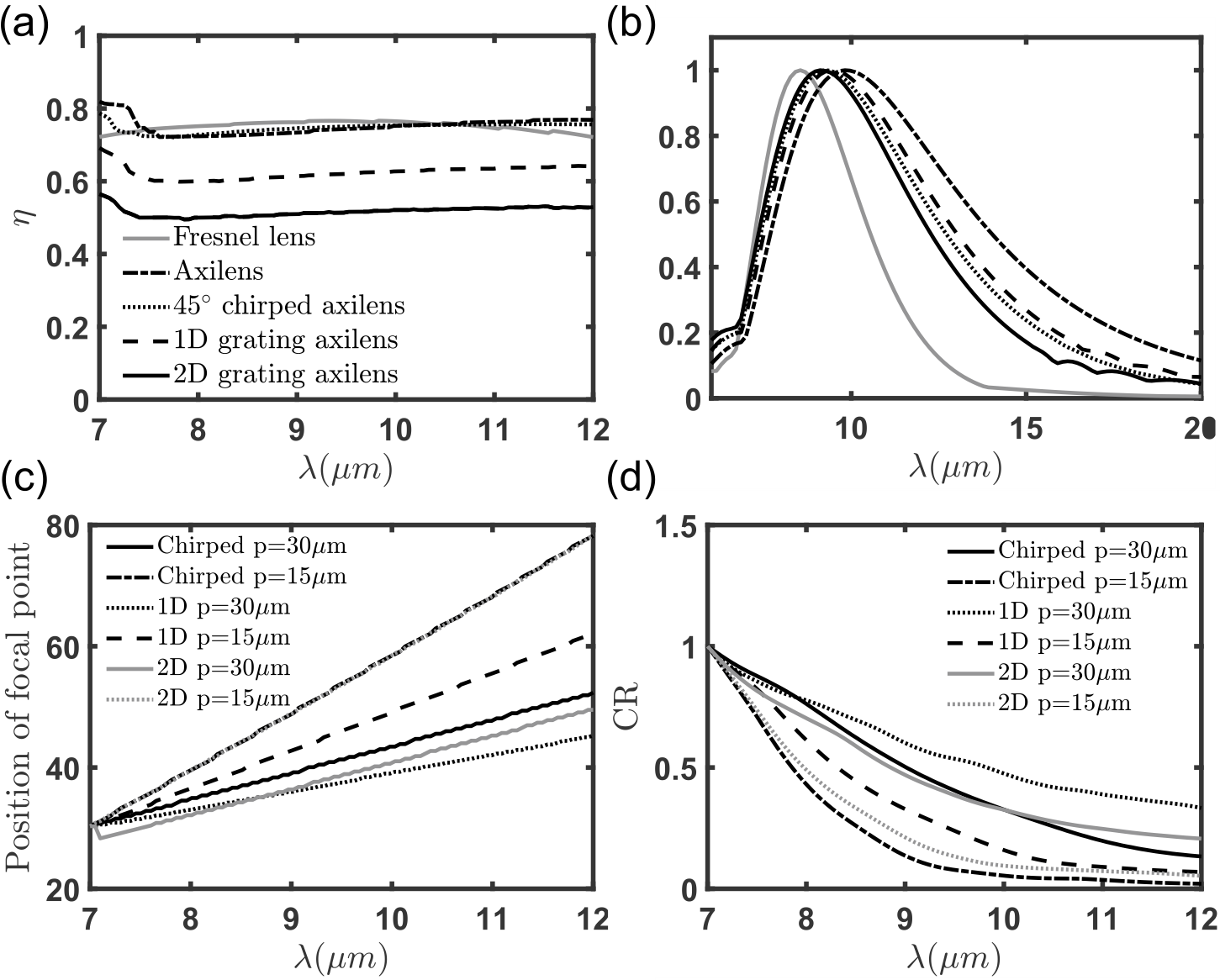}
\caption{Panels (a-b) show the focusing efficiency and the spectral response for the different structures specified in the legend. Panels (c-d) show the position of the focal spot with respect to the origin of the device and the cross talk ratio for the different modulated-phase axilenses specified in the legend, respectively.}
\label{Fig5}
\end{figure}

To further characterize the focusing behavior of phase-modulated axilenses, we compute the spectral dependence of $\eta$, spectral response (SR), position of the focal point, and crosstalk ratio (CR) for the different types of modulated axilenses. Figure \ref{Fig5} (a) compares the behavior of $\eta$ for a standard Fresnel lens with $f_0={300{\mu}m}$, an axilens with $f_0=250{\mu}m, {\Delta}f=120{\mu}m$, 1D and 2D periodically modulated axilenses, and a $45^\circ$ chirped axilens (both with $p=15{\mu}m$). This different $f_0$ value has been chosen in order to center the depth of focusing region in the plane $z=300{\mu}m$ for all the devices. Despite the decrease in $\eta$ for the modulated structures with respect to a standard axilens, which is expected because only the first diffraction order are considered, the modulated axilenses maintain high efficiency (${\sim}72\%$ for $45^\circ$ chirped, ${\sim}61\%$ for periodic modulation) over the broad wavelength range $7{\mu}m{\sim}12{\mu}m$. Figure \ref{Fig5} (b) shows the SR of the same devices as in panel (a), which demonstrates significantly increased spectral responses spanning across the long-wavelength IR band compared to a Fresnel lens due to the underlying axilens phase profile. We also investigate in Fig. \ref{Fig5} (c) the shift of the focal point position with respect to the center of the devices when varying the incident wavelength of radiation. Our findings show that the focal point position shifts linearly when increasing the incident radiation wavelength. Decreasing the periodicity of the phase modulation from $p=30{\mu}m$ to $p=15{\mu}m$ increases the slope of the wavelength shift. We note that by reducing the periodicity even further, the reported linear trend will break down and the wavelength scaling of the diffraction angle should be considered instead.

To assess the potential of modulated axilenses for IR photodetection we study the cross talk ratio (CR) between different wavelengths. The CR is defined as follows:
\begin{equation}\label{cross talk}
    CR={\int_{x1}^{x2}I_{ol}(x,\lambda_0,\lambda)dx}\bigg/{\int_{x1}^{x2}I(x,\lambda_0)dx}
\end{equation}
where $I_{ol}(x,\lambda_0,\lambda)$ is the 1D intensity cut through the center of the overlapping focusing regions between incident wavelength $\lambda_0$ and $\lambda$. $I(x,\lambda_0)$ is the intensity cut through the center of focal point with incident wavelength $\lambda_0$. The CR quantifies the cross talk between $\lambda$ and $\lambda_0$ at the same location. In Fig. \ref{Fig5} (d) we show the computed CR for different modulated axilenses by fixing $\lambda_0=7{\mu}m$ and sweeping $\lambda$ from $7{\mu}m$ to $12{\mu}m$. The phase modulation with the smallest periodicity features the smallest CR at the same wavelength since the focal points of $\lambda_0$ and $\lambda$ are further separated by the periodic modulation of the phase of the device on the focal plane, as demonstrated in Fig. \ref{Fig5} (c). Consistently with the larger focal shifts reported in Fig. \ref{Fig5} (c), the $45^\circ$ chirped axilens features smaller CR values compared with the 1D periodically modulated axilens at the same $\lambda$. However, it is important to note that the CR drops below $0.35$ when increasing $\lambda$ to $9.5{\mu}m$. This leads to a smaller minimum resolvable wavelength interval when using our devices for spectroscopic applications, as discussed in the next section.

\section{microspectrometers with modulated axilenses in the LWIR}
\begin{figure}[h!]
\centering
\includegraphics[width=\linewidth]{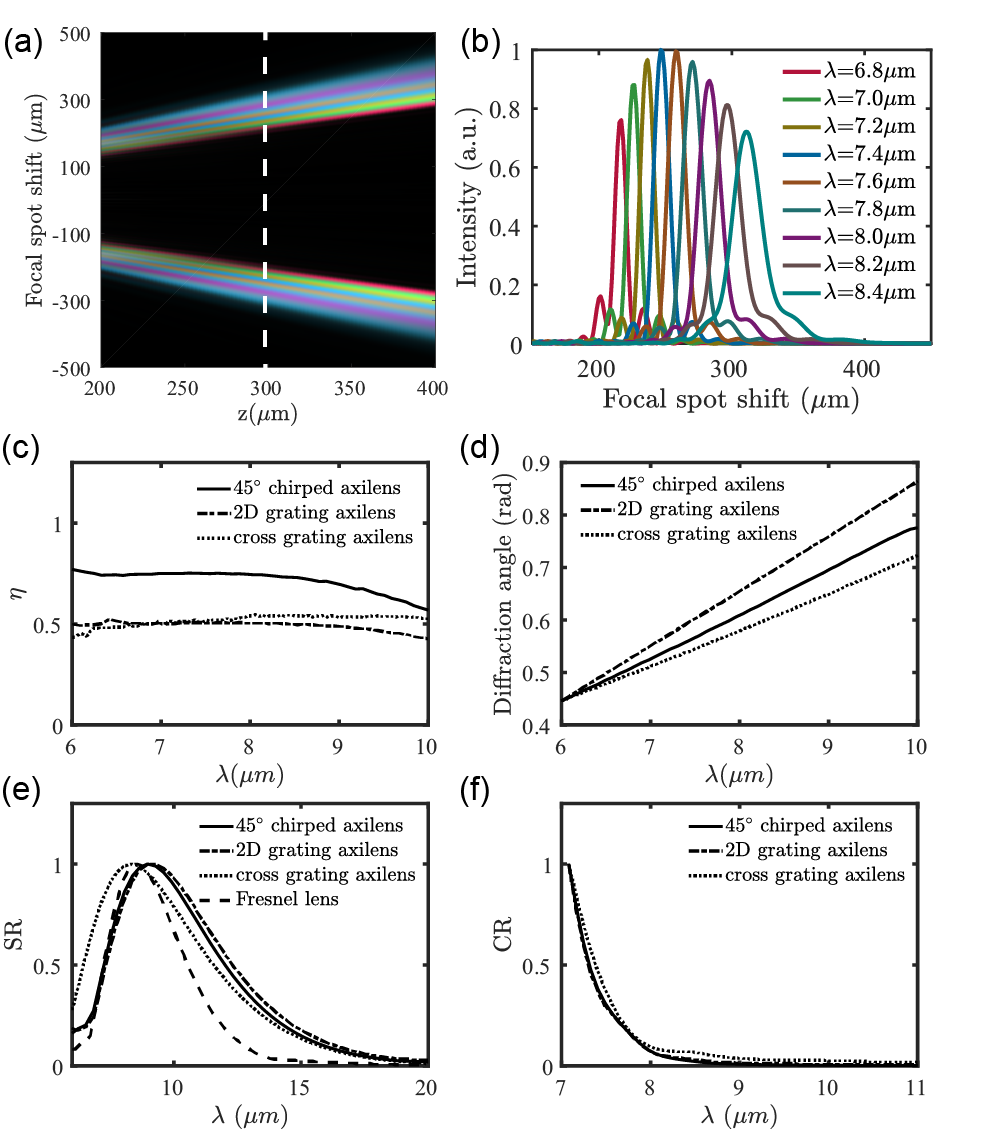}
  \caption{Panel (a) shows the spatial distribution of light with different wavelengths  (colors correspond to the legend in panel b) sampled over a $45^\circ$ plane for a 2D periodically modulated axilens. The axilens parameters are : $f_0=250{\mu}m$, ${\Delta}f=120{\mu}m$, $\lambda=9{\mu}m$, $D=100{\mu}m$, and $p=5{\mu}m$. (b) 1D normalized intensity profiles corresponding to different wavelengths at the focal distance $z=300{\mu}m$ (indicated by the white dashed line in panel a). Panels (c-d) show the wavelength dependent focusing efficiency $\eta$ and the diffraction angle of the focused spot for 2D modulated axilenses, respectively. Panel (e) displays the spectral response of 2D modulated axilenses compared with a Fresnel lens. Panel (f) displays the cross talk ratio of 2D modulated axilenses.}  
\label{Fig6}
\end{figure}

Building on the angular dispersion properties of 2D phase-modulated axilenses, we now demonstrate in Fig. \ref{Fig6} their potential for the engineering of single-lens compact microspectrometers. In particular, we show in Fig. \ref{Fig6} (a) the calculated intensity distribution sampled on a $45^\circ$ plane for different incident radiations with different wavelengths (and with the same intensity). We observe a clear spectral separation of the different incoming wavelengths at two focused locations on the same achromatic plane (indicated by the white dashed line). We used false colors to distinguish different wavelengths that correspond to the same colors shown in Fig. \ref{Fig6} (b). In Fig. \ref{Fig6} (b), we show the spatial distribution of the focused intensity corresponding to different wavelengths on the achromatic plane ($z=300{\mu}m$). We can clearly appreciate that different colors are diffracted into well-separated locations across this plane, enabling spectroscopic characteristic. 

We further characterize the wavelength dependence of $\eta$ and of the diffraction angles for the $45^\circ$ chirped axilens, the 2D periodically modulated axilens, and the cross grating axilens. The modulation period of these devices was fixed to $p=5{\mu}m$. Figure \ref{Fig6} (c) demonstrates that the three modulated axilenses achieve $\eta>50\%$ across a broad band in the LWIR. In Fig. \ref{Fig6} (d) we show the diffraction angle of the focal spots as a function of the incident wavelength, demonstrating the potential to engineer single lens focusing spectrometers with linear characteristics based on this technology. Figure \ref{Fig6} (e) shows the SR of a modulated axilens compared to a Fresnel lens with same size and with $f_0$=\,300\,${\mu}$m. We observe that the modulated axilens features a much broader spectral response across the LWIR. In Fig. \ref{Fig6} (f), we show the calculated cross talk ratio using Eq. \ref{cross talk} with $\lambda_0$\,=7\,${\mu}$m, which demonstrates better performance than the previous device with p=\,30\,$\mu$m and p=\,15\,$\mu$m, which were shown in Fig. \ref{Fig5} (d). In particular, we remark that the deflection angles shown in Fig. \ref{Fig6} (c) are linearly related to the incoming wavelength and the slope of this linear relation defines the angular dispersion $\frac{d\theta}{d\lambda}$ of the device.  For the devices considered here, we have selected diameter $D=100{\mu}m$. The angular dispersion for the 2D periodically modulated axilens is $\frac{d\theta}{d\lambda}=1*10^{-4}~rad/nm$. We also evaluate these parameters for the $45^\circ$ modulated and the cross grating modulated axilens that are $8.25*10^{-5}~rad/nm$ and $6.9*10^{-5}~rad/nm$, respectively. Therefore, 2D periodically modulated axilenses are more promising for spectroscopic applications. Remarkably, the angular dispersion value predicted for this miniaturized device is comparable to the one of a commercial spectrometer with a grating spatial frequency of $300~[lines/mm]$. This opens up exciting new scenarios for on-chip spectroscopic applications based on miniaturized single-lens components. 
Furthermore, we calculate the minimum resolvable wavelength interval of the 2D modulated axilens device using the following relation \cite{demtroder2014laser}:
\begin{equation}    
    \Delta\lambda=\frac{\lambda_c}{nD\frac{d\theta}{d\lambda}}
\end{equation}\label{minimum_interval}
where $\lambda_c$ is the wavelength in air, $D$ is the device diameter, n is the substrate material index. This yields, for our choice of parameters, $\Delta\lambda=240nm$ at $\lambda_c=8{\mu}m$. Notice that $\Delta\lambda$ is inversely related to the device aperture size $D$. Therefore, miniaturized modulated axilens-based spectrometers necessarily have limited $\Delta\lambda$ compared to large-scale commercial spectrometers with a typical aperture size of $10~cm$. However, the $\Delta\lambda$ can be further reduced by increasing device diameter D and/or further reducing the modulation periodicity p, depending on different application scenarios. For example, if we choose $D=200{\mu}m$ and $p=4{\mu}m$ our device can achieve $\Delta\lambda=96nm$, corresponding to $\sim 1\%$ of the central wavelength. In addition, phase apodization techniques can also be utilized to effectively reduce the aperture-driven diffraction effects \cite{goodman2005introduction}.

\section{microspectrometers with modulated axilenses in the MWIR}
\begin{figure}[t!]
\centering
\includegraphics[width=\linewidth]{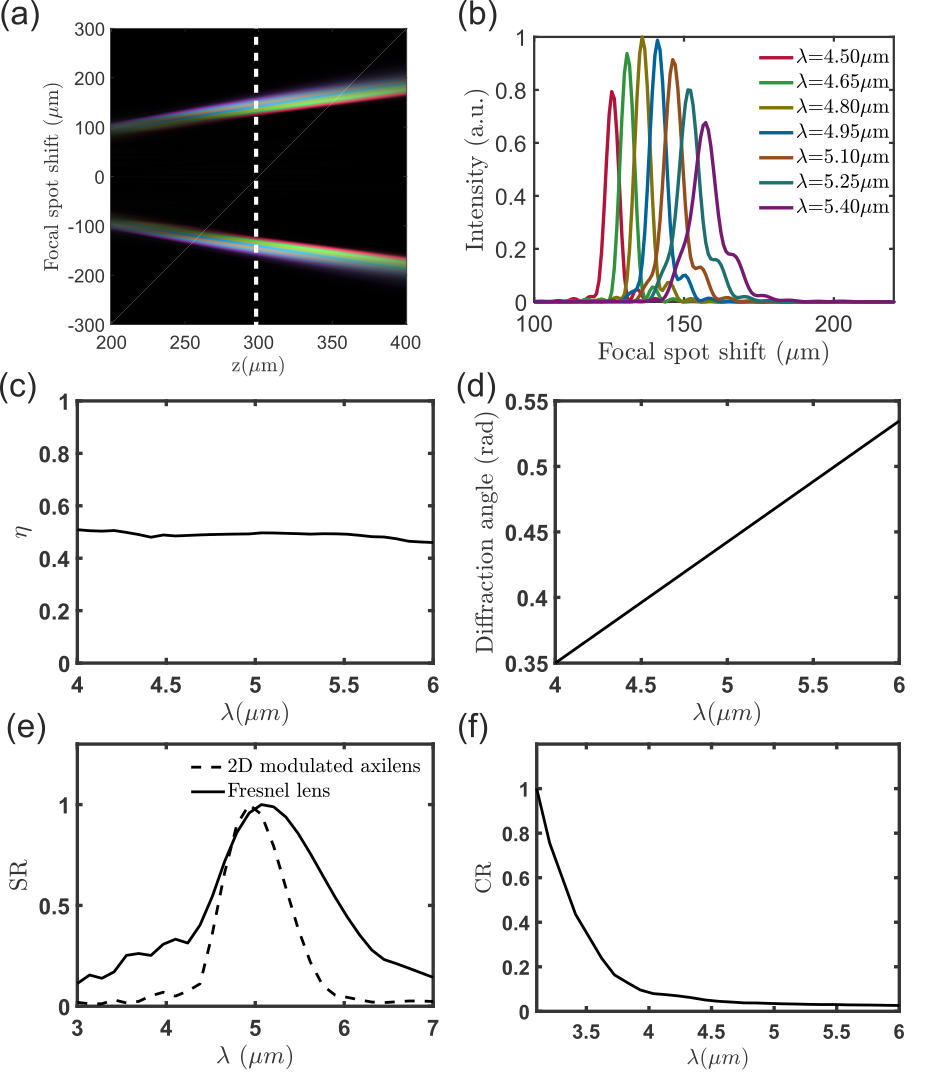}
  \caption{Panel (a) shows the spatial distribution of light with different wavelengths  (colors correspond to the legend in panel b) sampled over a $45^\circ$ plane for a 2D periodically modulated axilens. (b) 1D normalized intensity profiles corresponding to different wavelengths at the focal distance $z=300{\mu}m$ (indicated by the white dashed line in panel a). Panels (c-d) show the wavelength dependent focusing efficiency $\eta$ and the diffraction angle of the focused spot, respectively. Panels (e-f) display the spectral response and cross talk ratio of the 2D periodically modulated device in the MWIR, respectively.}  
\label{Fig7}
\end{figure}
Finally, we explore the applicability of our device concept to the MWIR wavelength range and we design the 2D modulated axilenses for optimal performance at $\lambda=5{\mu}m$. 
Our results are shown in Fig. \ref{Fig7} where devices are designed according to Eq. \ref{phase profile of axilens} with $\lambda=5~{\mu}m$. All the other device parameters of the device are the same as described in Fig. \ref{Fig6}. In particular, Fig. \ref{Fig7} (a) displays the diffracted 2D intensity distribution sampled on a $45^\circ$ plane of different radiation across the $4-6{\mu}m$ spectrum. We also show in  Fig. \ref{Fig7} (b) the 1D normalized intensity cuts corresponding to different incident wavelengths indicated by the white dashed line in Fig. \ref{Fig7} (a). A clear spectral separation of the different incoming wavelengths at two focused locations on the same achromatic plane is observed. Therefore, our device concept can also be utilized for focusing grating behavior across the MWIR spectral range.
Furthermore, we quantify its focusing efficiency and angular diffraction behavior at $z=300{\mu}m$ plane in Figs. \ref{Fig7} (c) and (d), respectively. The device still shows a focusing efficiency around $50\%$ over a broadband range ($4-6 {\mu}m$). Similar to the previous LWIR case, we now obtain an angular dispersion in Fig. \ref{Fig7} (d) as $\frac{d\theta}{d\lambda}=9*10^{-5}~rad/nm$. Therefore, the minimum resolvable wavelength interval for the MWIR implementation is $\Delta\lambda=165nm$ at $\lambda_c=5{\mu}m$. We further show the SR compared to a same-size Fresnel lens with $f_0=300{\mu}m$ in Fig. \ref{Fig7} (e). We observe that the modulated axilens features a much broader spectral response across the MWIR. Finally, in Fig. \ref{Fig7} (f), we show the calculated cross talk ratio using Eq. \ref{cross talk} with $\lambda_0$ =3 ${\mu}m$, which demonstrates even better performance compared to the LWIR implementations.

\section{Conclustion}
In conclusion, we have proposed and designed a compact 4-level phase-modulated axilens platform that can be directly integrated atop multi-band IR-FPA substrates to achieve LWIR and MWIR detection and  spectroscopic functionalities. We showed that these devices feature a programmable focusing behavior and broader spectral responses compared to standard Fresnel lenses and support an achromatic focusing plane. We further investigated different types of phase modulations and demonstrated the focusing of spectral components at different controllable locations on the same focal plane. Building on the concept of phase-modulated axilenses, we also demonstrate and characterize the spectroscopic behavior of these devices in the LWIR and MWIR. Specifically, we achieved minimum distinguishable wavelength intervals $\Delta\lambda=240nm$ and $\Delta\lambda=165nm$ at $\lambda_c=8{\mu}m$ and $\lambda_c=5{\mu}m$, respectively. The experimental characterization and testing of the proposed devices for operation at visible and near-infrared wavelengths is reported in reference \cite{britton2020phasemodulated}. The proposed compact devices add fundamental imaging and spectroscopic capabilities to traditional DOEs and are alternative solutions with respect to metasurface-based designs for applications to spectral sorting, multi-band IR imaging, photodetection, and spectroscopy.

\section*{Funding}{We acknowledge the support of the Army Research Laboratory under Cooperative Agreement Number W911NF-12-2-0023 for the development of theoretical and numerical models. Moreover we acknowledge funding from the NSF project: DMR 1709704 titled: \textit{Tunable Si-compatible Nonlinear Materials for Active Metaphotonics}.}

\providecommand{\newblock}{}

\end{document}